\documentclass[conference]{IEEEtran}
\IEEEoverridecommandlockouts
\usepackage{cite}
\usepackage{amsmath,amssymb,amsfonts}
\usepackage{graphicx}
\usepackage{textcomp}
\usepackage{xcolor}
\usepackage{acronym}
\usepackage{multicol}
\usepackage{subcaption}

\def\BibTeX{{\rm B\kern-.05em{\sc i\kern-.025em b}\kern-.08em
    T\kern-.1667em\lower.7ex\hbox{E}\kern-.125emX}}

\makeatletter
\newcommand{\mathleft}{\@fleqntrue\@mathmargin0pt}
\newcommand{\mathcenter}{\@fleqnfalse}
\makeatother

\DeclareMathOperator{\diag}{diag}
\DeclareMathOperator{\Tr}{Tr}
\DeclareMathOperator{\maximize}{maximize}
\DeclareMathOperator{\st}{subject~to}

\begin{document}

\title{Low Complexity Optimization for Line-of-Sight RIS-Aided Holographic Communications\\

\thanks{J. C. Ruiz-Sicilia and M. Di Renzo are with Universit\'e Paris-Saclay, CNRS, CentraleSup\'elec, Laboratoire des Signaux et Syst\'emes, 91192 Gif-sur-Yvette, France. (juan-carlos.ruiz-sicilia@centralesupelec.fr). 
M. Debbah is with Khalifa University of Science and Technology, Abu Dhabi, UAE.
H. V. Poor is with the Department of Electrical and Computer Engineering, Princeton University, Princeton, USA.
This work was supported in part by the European Commission through the H2020 MSCA 5GSmartFact project under grant agreement 956670, the H2020 ARIADNE project under grant agreement 871464, and H2020 RISE-6G project under grant agreement 101017011.}
}

\author{Juan Carlos Ruiz-Sicilia, Marco~Di~Renzo, Merouane Debbah, H. Vincent Poor}

\maketitle

\begin{abstract}
The synergy of metasurface-based holographic surfaces (HoloS) and reconfigurable intelligent surfaces (RIS) is considered a key aspect  for future communication networks. However, the optimization of dynamic metasurfaces requires the use of numerical algorithms, for example, based on the  singular value decomposition (SVD) and gradient descent methods, which are usually computationally intensive, especially when the number of elements is large. In this paper, we analyze low complexity designs for RIS-aided HoloS communication systems, in which the configurations of the HoloS transmitter and the RIS are given in a closed-form expression. We consider implementations based on diagonal and non-diagonal RISs. Over line-of-sight channels, we show that the proposed schemes provide performance that is close to that offered by complex numerical methods.
\end{abstract}


\section{Introduction}

The paradigm of smart radio environment (SRE) has gained considerable attention in the wireless research community \cite{Gacanin,Renzo2019,Huang2020}. This new concept refers to the use of intelligent (metamaterial) surfaces to manipulate the electromagnetic waves with high flexibility, leading to higher bit rates and a larger number of supported devices compared with current wireless networks \cite{Dardari21}. The SRE paradigm is based on two technologies: the holographic surface (HoloS) and the reconfigurable intelligent surface (RIS). A HoloS is an active dynamic metasurface that is employed as a flexible antenna. An RIS is a nearly-passive dynamic metasurface capable of shaping the electromagnetic waves \cite{DBLP:journals/jsac/RenzoZDAYRT20}, \cite{Wu2021}. For example, the waves that reach an RIS can be intelligently reflected to optimize the end-to-end channel.

Despite the potential benefits of utilizing dynamic metasurfaces in wireless communications, the optimization of RIS-aided HoloS systems is an open problem. In \cite{Perovic}, the authors optimize the transmitter and the RIS iteratively, by using a projected gradient method. In \cite{Bartoli2022}, a simplified method to optimize an RIS-aided system is presented. The approach is based on the concept of focusing function for optimizing the RIS \cite{Miller}. Also, the transmitter is optimized by computing the singular value decomposition of the channels and the water-filling power allocation \cite{Tse}. In \cite{ruiz2023}, the authors analyze different strategies to optimize RIS-aided HoloS systems as a function of the channel state information. 

In general, the schemes reported in the literature rely on complex numerical algorithms, whose scalability is compromised when the size of the surfaces becomes too large. In this paper, we show that, in line-of-sight channels, the optimal designs of RISs and HoloSs can be formulated in closed-form expressions, which are simple to compute. Also, the obtained analytical formulations give insights on the achievable system performance and how it depends on, e.g., the size of the surfaces and the geometry of the setup.

The rest of the present paper is organized as follows. In Section \ref{sec:SystemModel}, we present the system model. In Section \ref{sec:CaseStudies}, we introduce the proposed schemes. In Section \ref{sec:NumResults}, we illustrate numerical results and compare the performance of the considered schemes. Finally, conclusions are given in Section \ref{sec:Conclusions}.

\textit{Notation}: Bold lower and upper case letters represent vectors and matrices. $\mathbb{C}^{a \times b}$ denotes the space of complex matrices of dimensions $a \times b$. $(\cdot)^\dag$ denotes the Hermitian transpose. $\mathbf{I}_N$ denotes the $N \times N$ identity matrix. $\diag(\mathbf{x})$ denotes a square diagonal matrix whose diagonal elements are $\mathbf{x}$. $\Tr(\mathbf{X})$ is the trace of matrix $\mathbf{X}$, and $\mathbb{E}\{\cdot\}$ is the expectation operator. The notation $\mathbf{A}\succeq(\succ) \mathbf{B}$ means that $\mathbf{A}-\mathbf{B}$ is positive semidefinite (definite). $\mathbf{A}(i,k)$ denotes the $k$-th element of the $i$-th row of matrix $\mathbf{A}$. $\mathbf{a}_k$ denotes the $k$-th column of matrix $\mathbf{A}$. $j$ is the imaginary unit. $\mathcal{CN}(\mu,\sigma^2)$ denotes the complex Gaussian distribution with mean $\mu$ and variance $\sigma^2$.

\section{System model}
\label{sec:SystemModel}
\begin{figure}[!t]
    \centering
\includegraphics[width=\columnwidth]{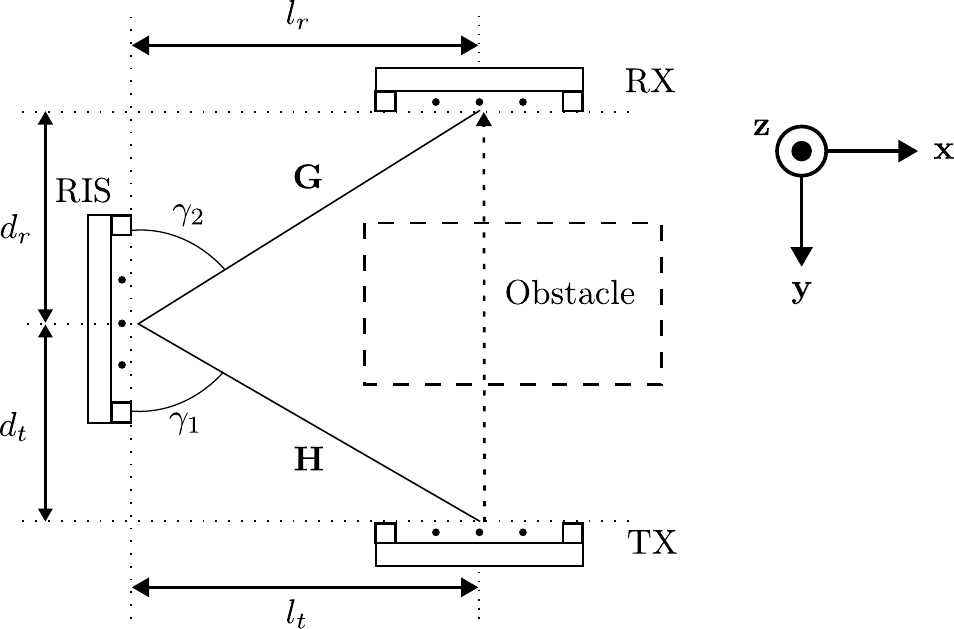}
    \caption{Top view of the considered system model.}
    \label{fig:system_model} \vspace{-0.5cm}
\end{figure}

Consider the RIS-aided HoloS communication system sketched in Fig. \ref{fig:system_model}. The transmit and receive surfaces are deployed parallel to the $xz$ plane, and the RIS lies on a wall that is parallel the $yz$ plane. The distances from the plane that contains the transmitter and from the plane that contains the receiver to the center of the RIS are $d_t$ and $d_r$, respectively. The distance between the center of the transmit HoloS and the plane containing the RIS is $l_t$, and the distance between the midpoint of the receive HoloS and the plane containing the RIS is $l_r$. The three surfaces are deployed at the same height. For simplicity, we assume that the direct link between the transmitter and the receiver is blocked by an obstacle. 

Each HoloS is modeled as a uniform rectangular array (URA). The transmitter and receiver are equipped with $L= L_x L_z$ and $M = M_x M_z$ antenna elements, respectively, where $L_x$ ($M_x$) and $L_z$ ($M_z$) are the number of antenna elements on the $x$-axis and $z$-axis. The RIS consists of $N = N_y N_z$ unit cells, where $N_y$ and $N_z$ are the number of antenna elements on the $y$-axis and $z$-axis. In all surfaces, the separation between the centers of two adjacent elements is $\delta = \lambda/2$, where $\lambda$ is the wavelength, in order to avoid the mutual coupling. 

The positions of the $l$-th transmit element, $m$-th receive element, and $n$-th unit cell of the RIS are denoted by $\mathbf{r}^l_t = (x^l_t,y^l_t,z^l_t)$, $\mathbf{r}^m_r = (x^m_r,y^m_r,z^m_r)$ and $\mathbf{r}^n_\mathrm{ris} = (x^n_\mathrm{ris},y^n_\mathrm{ris},z^n_\mathrm{ris} )$, respectively. They can be formulated as
\mathleft
\begin{equation}
\label{eq:paramTx}
    \mathbf{r}^l_t = \left(l_t + \delta l_x - \frac{\delta}{2}(L_x+1), d_{t},\delta l_z - \frac{\delta}{2}(L_z+1) \right)
\end{equation}
\begin{multline}
    \mathbf{r}^m_r =\\ \left(\vphantom{\frac12} l_r + \delta m_x - \frac{\delta}{2}(M_x+1), -d_{r},
    \delta m_z - \frac{\delta}{2}(M_z+1) \right)
\end{multline}
\begin{equation}
\label{eq:paramRIS}
    \mathbf{r}^n_\mathrm{RIS} = \left( 0, n_yd-\frac{\delta}{2}(N_{y}+1), n_z \delta-\frac{\delta}{2}(N_{z}+1) \right)
\end{equation}
\mathcenter
where $l = (l_z-1)L_z+l_y$, $m = (m_z-1)M_z+m_y$ and $n= (n_x-1)N_x+n_y$, and $l_z$ ($m_z$), with $l_y$ ($m_y$) denoting the indices of the URAs along the $z$-axis and $y$-axis, respectively. Similarly, $n_x$ and $n_y$ denote the indices of the unit cells of the RIS along the $x$-axis and $y$-axis, respectively.

The RIS is modeled as a matrix $\mathbf{\Phi}$ that contains $N \times N$ reflection coefficients. Considering that the RIS is nearly-passive, the matrix needs to fulfill the condition $\mathbf{\Phi} \mathbf{\Phi}^\dag = \mathbf{I}_N$. We refer to this general RIS as a non-diagonal RIS. If $\mathbf{\Phi}$ is a diagonal matrix with unit-modulus coefficients, the RIS is referred to as a diagonal RIS \cite{Bartoli2022}. 

The channels from the transmitter to the RIS and from the RIS to the receiver are denoted as $\mathbf{H} \in \mathbb{C}^{N \times L}$ and $\mathbf{G} \in \mathbb{C}^{M \times N}$, respectively. The channel between the $l$-th radiating element of the transmitter and the $n$-th element of the RIS is expressed as
\begin{equation}
    \mathbf{H}(n,l) = \frac{e^{j k_0 d_1(n,l)}}{4\pi d_1(n,l)}
\end{equation}
where $k_0 = 2 \pi/\lambda$ and $d_1(n,l)$ is the distance between the two elements. Similarly, the channel from the $n$-th element of the RIS to the $m$-th receiving antenna is formulated as
\begin{equation}
    \mathbf{G}(m,n) = \frac{e^{j k_0 d_2(m,n)}}{4\pi d_2(m,n)}
\end{equation}
where $d_2(m,n)$ is the distance between the two elements. 

Hence, the received signal $\mathbf{y} \in \mathbb{C}^{M}$ is given by
\begin{equation}
    \mathbf{y} = \mathbf{G}\mathbf{\Phi}\mathbf{H} \mathbf{x} + \mathbf{n}
\end{equation}
where $\mathbf{n}$ is the additive white Gaussian noise with $\mathbf{n} \sim \mathcal{CN}(0, \sigma^2\mathbf{I}_N)$ and $\mathbf{x} \in \mathbb{C}^{L}$ is the transmitted signal. The end-to-end channel is denoted as $\mathbf{Z}(\mathbf{\Phi}) = \mathbf{G}\mathbf{\Phi}\mathbf{H}$. The symbols in $\mathbf{x}$ are distributed according to a circularly symmetric complex Gaussian distribution with $\mathbb{E}\{ \mathbf{x} \mathbf{x}^\dag \} = \mathbf{Q}$, where $\mathbf{Q}$ is the covariance matrix. The matrix $\mathbf{Q}$ needs to fulfill the constraint $\Tr(\mathbf{Q}) \leq P_T$, where $P_T$ is the transmit power budget.

Then, the achievable rate can be computed as
\begin{equation}
\label{eq:capacity}
    R(\boldsymbol{\Phi}, \mathbf{Q}) = \log_2\left|\mathbf{I}_M + \frac{ (\mathbf{G}\boldsymbol{\Phi}\mathbf{H}) \mathbf{Q} (\mathbf{G}\boldsymbol{\Phi}\mathbf{H})^\dag }{\sigma^2 } \right|
\end{equation}

\section{Considered Schemes}
\label{sec:CaseStudies}

This section presents four schemes that maximize the rate in \eqref{eq:capacity}, by optimizing the RIS configuration and the transmit covariance matrix, i.e., the matrices $\mathbf{\Phi}$ and $\mathbf{Q}$. 

\subsection{Non-Diagonal RIS Based on Numerical Methods}
\label{subsec:ND_num}
In this case study, we consider a non-diagonal RIS and we optimize  $\mathbf{\Phi}$ and $\mathbf{Q}$ by using numerical methods. This case study constitutes, therefore, the benchmark scheme \cite{Bartoli2022}.

The optimization problem can be formulated as follows:
\begin{subequations}\label{eq:capacity_gen_prob}
\begin{align}
\underset{\boldsymbol{\Phi},\mathbf{Q}}{\maximize}
& \ R(\boldsymbol{\Phi}, \mathbf{Q}) \label{eq:opt_problem}\\
\st & \ \Tr(\mathbf{Q})\le
P_{t}; \mathbf{Q}\succeq\mathbf{0};\\
 & \ \mathbf{\Phi}\mathbf{\Phi}^\dag = \mathbf{I}_N
\end{align}
\end{subequations}

By applying the singular value decomposition (SVD) to the channel matrices $\mathbf{H}$ and $\mathbf{G}$, we obtain
\begin{equation}
    \mathbf{H} = \mathbf{U}_\mathrm{H} \mathbf{S}_\mathrm{H} \mathbf{V}_\mathrm{H}^\dag, \qquad
    \mathbf{G} = \mathbf{U}_\mathrm{G} \mathbf{S}_\mathrm{G} \mathbf{V}_\mathrm{G}^\dag
\end{equation}
where $\mathbf{U}_\mathrm{H} \in \mathbb{C}^{N \times N}$, $\mathbf{V}_\mathrm{H} \in \mathbb{C}^{L \times L}$, $\mathbf{U}_\mathrm{G} \in \mathbb{C}^{M \times M}$ and
$\mathbf{V}_\mathrm{G} \in \mathbb{C}^{N \times N}$ are unitary matrices, and $\mathbf{S}_\mathrm{H}$ and $\mathbf{S}_\mathrm{G}$ are given by
\begin{align}
    \mathbf{S}_\mathrm{H} &= \diag \{ \lambda_{\mathrm{H},1}, \lambda_{\mathrm{H},2},..., \lambda_{\mathrm{H},K_1}\} \; \in \; \mathbb{C}^{N \times L}\\
    \mathbf{S}_\mathrm{G} &= \diag \{ \lambda_{\mathrm{G},1}, \lambda_{\mathrm{G},2},..., \lambda_{\mathrm{G},K_2}\}\; \in \; \mathbb{C}^{M \times N}
\end{align}
where $K_1 = \min (L,N)$, $K_2 = \min (M,N)$, and $\lambda_{\mathrm{H},k}$ and $\lambda_{\mathrm{G},k}$ are the $k$-th largest singular values of $\mathbf{H}$ and $\mathbf{G}$. 

Hence, the end-to-end channel $\mathbf{Z}(\mathbf{\Phi})$ can be written as
\begin{equation}
\label{eq:channelDecomp}
    \mathbf{Z}(\mathbf{\Phi}) = \mathbf{U}_\mathrm{G} \mathbf{S}_\mathrm{G} \mathbf{V}_\mathrm{G}^\dag \mathbf{\Phi}\mathbf{U}_\mathrm{H} \mathbf{S}_\mathrm{H} \mathbf{V}_\mathrm{H}^\dag
\end{equation}

According to \cite[Proposition 1]{Bartoli2022}, the solution of the optimization problem in \eqref{eq:capacity_gen_prob} is given by 
\begin{equation}
\label{eq:RIS_ND_NUM}
    \mathbf{\Phi}^\mathrm{ND-NUM} = \mathbf{V}_\mathrm{G} \mathbf{U}_\mathrm{H}^\dag
\end{equation}
\begin{equation}
\label{eq:Q_ND_NUM}
    \mathbf{Q}^\mathrm{ND-NUM} = \mathbf{V}_\mathrm{H} \mathbf{P}^\mathrm{ND-NUM} \mathbf{V}_\mathrm{H}^\dag
\end{equation}
where $\mathbf{P}^\mathrm{ND-NUM}$ is a diagonal matrix containing the power allocated to each communication mode. 

The methodology to compute $\mathbf{P}^\mathrm{ND-NUM}$ is presented in Subsection \ref{subsec:powerAlloc}, where the signal-to-noise ratio (SNR) of the $k$-th communication mode is defined as
\begin{equation}
    s_k^\mathrm{ND-NUM} = \frac{|\lambda_{\mathrm{H},k}|^2 |\lambda_{\mathrm{G},k}|^2}{\sigma^2}
\end{equation}

\subsection{Non-Diagonal RIS Based on Analysis}
\label{sub:AnaND}
The scheme summarized in the previous subsection requires the numerical computation of the precoding and decoding matrices through an SVD, which is the typical approach adopted in the literature. In this sub-section, we present an approximated closed-form expression for the optimal $\boldsymbol{\Phi}$ and $\mathbf{Q}$ that are solutions of the problem in \eqref{eq:capacity_gen_prob}.

To this end, we introduce the matrices $\mathbf{U}_\mathrm{G}^\mathrm{PSWF}$, $\mathbf{V}_\mathrm{G}^\mathrm{PSWF}$, $\mathbf{U}_\mathrm{H}^\mathrm{PSWF}$ and $\mathbf{V}_\mathrm{H}^\mathrm{PSWF}$, which are defined, similar to \cite{Bartoli2022}, as 
\begin{gather}
\mathbf{V}_\mathrm{H}^\mathrm{PSWF} = \mathbf{F}_\mathrm{Tx,RIS} \mathbf{N}_\mathrm{Tx,RIS}\\
\mathbf{U}_\mathrm{H}^\mathrm{PSWF} = \mathbf{F}_\mathrm{RIS,Tx} \mathbf{N}_\mathrm{RIS,Tx} \label{eq:U_H}\\
\mathbf{V}_\mathrm{G}^\mathrm{PSWF} = \mathbf{F}_\mathrm{RIS,Rx} \mathbf{N}_\mathrm{RIS, Rx} \label{eq:V_G}\\
\mathbf{U}_\mathrm{G}^\mathrm{PSWF} = \mathbf{F}_\mathrm{Rx,RIS} \mathbf{N}_\mathrm{Rx,RIS}
\end{gather}
where $\mathbf{F}_\mathrm{Tx,RIS} \in \mathbb{C}^{L\times L}$, $\mathbf{F}_\mathrm{RIS,Tx} \in \mathbb{C}^{N\times N}$, 
$\mathbf{F}_\mathrm{RIS,Rx} \in \mathbb{C}^{N\times N}$ and
$\mathbf{F}_\mathrm{Rx,RIS} \in \mathbb{C}^{M\times M}$ are unitary diagonal matrices, and $\mathbf{N}_\mathrm{Tx,RIS} \in \mathbb{C}^{L\times L}$, $\mathbf{N}_\mathrm{RIS,Tx} \in \mathbb{C}^{N\times N}$, 
$\mathbf{N}_\mathrm{RIS,Rx} \in \mathbb{C}^{N\times N}$ and
$\mathbf{N}_\mathrm{Rx,RIS} \in \mathbb{C}^{M\times M}$ are unitary non-diagonal matrices.

The matrices $\mathbf{F}_\mathrm{A,B}$ for $\mathrm{A,B} = \{\mathrm{Tx},\mathrm{Rx},\mathrm{RIS}\}$ are referred to as focusing functions \cite{Miller}. The matrices $\mathbf{N}_\mathrm{A,B}$ determine, on the other hand, the spatial multiplexing capabilities of the scheme. In \cite{Bartoli2022}, the authors have analyzed the case study in which the surfaces $\mathrm{A}$ and $\mathrm{B}$ are parallel to each other. In the present paper, we consider the setup in which the two surfaces are deployed as depicted in Fig. 1, i.e., they are orthogonal to each other. The details of the derivation are available in companion journal version of the present paper.

Similar to \cite{Bartoli2022}, the matrices $\mathbf{F}_\mathrm{Tx,RIS}$, $\mathbf{F}_\mathrm{RIS,TX}$ and $\mathbf{F}_\mathrm{RIS,RX}$ are focusing functions, which can be expressed as follows:
\begin{multline}
\mathbf{F}_\mathrm{Tx,RIS} = \diag \{e^{-jk_0 d_\mathrm{Tx,RIS}(1)},\\ e^{-jk_0 d_\mathrm{Tx,RIS}(2)},...,e^{-jk_0 d_\mathrm{Tx,RIS}(L)} \}
\end{multline}
\begin{multline}
\mathbf{F}_\mathrm{RIS,Tx} = \diag \{e^{jk_0 d_\mathrm{RIS,Tx}(1)},\\ e^{jk_0 d_\mathrm{RIS,Tx}(2)},...,e^{jk_0 d_\mathrm{RIS,Tx}(N)} \}
\end{multline}
\begin{multline}
\mathbf{F}_\mathrm{RIS,Rx} = \diag \{e^{-jk_0 d_\mathrm{RIS,Rx}(1)},\\ e^{-jk_0 d_\mathrm{RIS,Rx}(2)},...,e^{-jk_0 d_\mathrm{RIS,Rx}(N)} \}
\end{multline}
where $d_\mathrm{Tx,RIS}(l)$ is the distance from the $l$-th element of the transmitter to the center of the RIS, and $d_\mathrm{RIS,Tx}(n)$ and $d_\mathrm{RIS, Rx}(n)$ are the distances from the $n$-th element of the RIS to the center of the transmitter and receiver, respectively.

Also, the columns of the matrix $\mathbf{N}_\mathrm{A,B}$ correspond to sampled versions of the basis functions that are used at $\mathrm A$. Let $\{\varphi_k(x, z)\}$, $\{\psi_k(y, z)\}$ and $\{\phi_k(x, z)\}$ be the set of continuous basis functions that correspond to the matrices $\mathbf{N}_\mathrm{Tx,RIS}$, $\mathbf{N}_\mathrm{RIS,Tx}$ and $\mathbf{N}_\mathrm{RIS, Rx}$, respectively. Let us assume that the following conditions hold:
\begin{equation}
\label{eq:largeShift1}
    r_1 \gg \Delta x_T, \Delta z_T, \Delta y_\mathrm{RIS}, \Delta z_\mathrm{RIS}
\end{equation}
where $r_1 = \sqrt{l_t^2 + d_t^2}$, $2\Delta x_T = \delta L_x$, $2\Delta z_T = \delta L_z$, $2\Delta y_\mathrm{RIS} = \delta N_y$ and $2\Delta z_\mathrm{RIS} = \delta N_z$. Then, the basis functions of the link between the transmitter and the RIS are approximately
\begin{align}
\varphi_k(x, z) &= 
\alpha_{k_x}\left(\frac{x - l_t}{\Delta x_T}, c_{xy}^{(1)}\right) 
\alpha_{k_z}\left(\frac{z}{\Delta z_T}, c_{zz}^{(1)}\right) \label{eq:modeContTx}\\
\psi_k(y, z) &=
\alpha_{k_y}\left(\frac{y}{\Delta y_\mathrm{RIS}}, c_{xy}^{(1)} \right)
\alpha_{k_z} \left(\frac{z}{\Delta z_\mathrm{RIS}}, c_{zz}^{(1)}\right)\label{eq:modeContRIS}
\end{align}
where $\alpha_{k_u}(u,c)$ is the $k_u$-th prolate spheroidal wave function (PSWF) with bandwidth parameter $c$ in the Flammer notation \cite{flammer}. The parameters $c_{xy}^{(1)}$, and $c_{zz}^{(1)}$ depend on the network geometry. Considering the setup depicted in Fig. 1, we have
\begin{equation}
    c_{xy}^{(1)} = \frac{\Delta x_T \Delta y_\mathrm{RIS} k_0}{ r_1} \frac{\sin 2\gamma_1}{2}\quad
    c_{zz}^{(1)} = \frac{\Delta z_T \Delta z_\mathrm{RIS}k_0}{ r_1}
\end{equation}
where $\gamma_1 = \sin^{-1} l_t/r_1$ is the angle depicted in Fig. 1. In practice, the PSWFs can be approximately computed by utilizing the Legendre polynomial expansion. To obtain the results illustrated in Section IV, we have used the implementation provided in \cite{PSWFgen}.

The functions $\alpha_{k_x}(x,c_x)$ and $\alpha_{k_z}(z,c_z)$ correspond to the eigenvalues $\nu_{k_x}$ and $\nu_{k_z}$, respectively. The coupling intensity of the $k$-th communication mode supported by the RIS-aided channel is $|\nu_{k}|^2 =|\nu_{k_x}|^2|\nu_{k_z}|^2$ for any $k_x$ and $k_z$. As usual practice, we assume that the basis functions are ordered in a decreasing order as a function of $|\nu_{k}|^2$. Although any combination of $k_x$ and $k_z$ results in a feasible basis function, only the communication modes whose coupling intensity $|\nu_{k}|^2$ is sufficiently high are relevant for data transmission. Specifically, the number of highly-coupled modes, i.e., the degrees of freedom (DoF), of the considered system model is
\begin{equation}
    N_1 \approx \frac{S_\mathrm{Tx} S_\mathrm{RIS}}{\lambda^2 r_1^2}\frac{\sin 2\gamma_1}{2}
\end{equation}
where $S_\mathrm{Tx} = (2\Delta x_T)(2\Delta z_T)$ is the area of the transmit HoloS and $S_\mathrm{RIS} = (2 \Delta y_\mathrm{RIS})(2 \Delta z_\mathrm{RIS})$ is the area of the RIS. 

As far as the link from the RIS to the receiver is concerned, a similar line of thought can be applied. Specifically, let us assume that the following conditions hold:
\begin{equation}
\label{eq:largeShift2}
    r_2 \gg \Delta x_R, \Delta z_R, \Delta y_\mathrm{RIS}, \Delta z_\mathrm{RIS}
\end{equation}
where $r_2 = \sqrt{l_r^2 + d_r^2}$, $2\Delta x_R = \delta M_x$ and $2\Delta z_R = \delta M_z$. Then, the optimal basis functions at the RIS are
\begin{equation}
    \phi_k(y, z) =
\alpha_{k_y}\left(\frac{y}{\Delta y_\mathrm{RIS}}, c_{xy}^{(2)} \right)
\alpha_{k_z} \left(\frac{z}{\Delta z_\mathrm{RIS}}, c_{zz}^{(2)}\right)
\end{equation}
where the parameters $c_{xy}^{(2)}$ and $c_{zz}^{(2)}$ are
\begin{equation}
    c_{xy}^{(2)} = \frac{\Delta x_T \Delta y_\mathrm{RIS} k_0}{ r_2} \frac{\sin 2\gamma_2}{2} \quad
    c_{zz}^{(2)} = \frac{\Delta z_T \Delta z_\mathrm{RIS} k_0}{ r_2}
\end{equation}
with $\gamma_2 = \sin^{-1} l_r/r_2$ defined in Fig. 1. 

Similar to the transmitter-RIS link, the $k$-th basis function $\phi_k(y, z)$ corresponds to the $k$-th largest coupling intensity coefficient $|\mu_k|^2$. Thus, the number of DoF of the RIS-receiver link is
\begin{equation}
    N_2 \approx \frac{S_\mathrm{Rx} S_\mathrm{RIS}}{\lambda^2 r_2^2}\frac{\sin 2\gamma_2}{2}
\end{equation}
where $S_\mathrm{Rx} = (2\Delta x_R)(2\Delta z_R)$ is the area of the receiving surface.

The columns of $\mathbf{N}_\mathrm{Tx,RIS}$, $\mathbf{N}_\mathrm{RIS,Tx}$ and $\mathbf{N}_\mathrm{RIS,Rx}$ are computed by sampling the continuous basis functions according to the positions of the radiating elements. Also, the eigenvalues must be scaled after the sampling. In particular, the continuous eigenvalues for both links are $|\lambda_{\mathrm{H},k}^\mathrm{PSWF}|^2 = |\nu_k \delta^2|^2$ and $ |\lambda_{\mathrm{G},k}^\mathrm{PSWF}|^2 = |\mu_k \delta^2|^2$.

Based on the obtained precoding and decoding matrices, the optimal RIS configuration is obtained as follows:
\begin{equation}
\label{eq:RIS_ND_PSWF}
    \mathbf{\Phi}^\mathrm{ND-PSWF} = \mathbf{V}_\mathrm{G}^\mathrm{PSWF} \left(\mathbf{U}_\mathrm{H}^\mathrm{PSWF} \right)^\dag
\end{equation}
and the transmit covariance matrix is
\begin{equation}
\label{eq:Q_PSWF}
    \mathbf{Q}^\mathrm{PSWF} = \mathbf{V}_\mathrm{H}^\mathrm{PSWF} \mathbf{P}^\mathrm{PSWF} \left(\mathbf{V}_\mathrm{H}^\mathrm{PSWF} \right)^\dag
\end{equation}

Also, $\mathbf{P}^\mathrm{PSWF}$ is the power allocation matrix defined in Subsection \ref{subsec:powerAlloc}, where the SNR of the $k$-th communication mode is given by \begin{equation}
    s_k^\mathrm{ND-PSWF} = \frac{|\lambda_{\mathrm{H},k}^\mathrm{PSWF}|^2 |\lambda_{\mathrm{G},k}^\mathrm{PSWF}|^2}{\sigma^2}
\end{equation}

\subsection{Diagonal RIS Based on Numerical Methods}
\label{subsec:foc_num}
To reduce the implementation complexity, we analyze the typical case study in which the RIS is characterized by a diagonal matrix. The corresponding optimization problem can be formulated as follows:
\begin{subequations}
\begin{align}
\underset{\boldsymbol{\Phi},\mathbf{Q}}{\maximize}
& \ R(\boldsymbol{\Phi}, \mathbf{Q})\\
\st & \ \Tr(\mathbf{Q})\le
P_{t};\mathbf{Q}\succeq\mathbf{0};\\
 & \ \mathbf{\Phi}\mathbf{\Phi}^\dag = \mathbf{I}_N; \mathbf{\Phi}(n,k)=0 \quad \forall \; n \not= k
\end{align}
\end{subequations}

In general, the formulated problem is non-convex and requires computationally intensive numerical algorithms to be solved. However, a closed-form sub-optimal solution was recently proposed in \cite{Bartoli2022} under the assumption that the surfaces are parallel to each other. By using a similar line of thought as that in \cite{Bartoli2022}, an approximated expression of $\mathbf{\Phi}$ for the system model in Fig. 1 is
\begin{equation}
\label{eq:RIS_FOC}
    \mathbf{\Phi}^\mathrm{FOC} = \mathbf{F}_\mathrm{RIS,Rx} \mathbf{F}_\mathrm{RIS,Tx}^\dag
\end{equation}

Inserting \eqref{eq:U_H}, \eqref{eq:V_G}, and \eqref{eq:RIS_FOC} into \eqref{eq:channelDecomp}, the end-to-end channel can be written as follows:
\begin{equation}
    \mathbf{Z}(\mathbf{\Phi}^\mathrm{FOC}) = \mathbf{U}_\mathrm{G} \mathbf{S}_\mathrm{G} \mathbf{N}_\mathrm{RIS, Rx}^\dag \mathbf{N}_\mathrm{RIS,Tx} \mathbf{S}_\mathrm{H} \mathbf{V}_\mathrm{H}^\dag
\end{equation}
since $\mathbf{F}_\mathrm{RIS,Rx}^\dag  \mathbf{F}_\mathrm{RIS,Rx} = \mathbf{I}_N$ and $\mathbf{F}_\mathrm{RIS,Tx}^\dag \mathbf{F}_\mathrm{RIS,Tx} = \mathbf{I}_N$. 

The approximation for $\mathbf{\Phi}^\mathrm{FOC}$ in \eqref{eq:RIS_FOC} is optimal only when $\mathbf{N}_\mathrm{RIS, Rx}^\dag \mathbf{N}_\mathrm{RIS,Tx} = \mathbf{I}_N$, since the end-to-end channel $\mathbf{Z}(\mathbf{\Phi}^\mathrm{FOC})$ is diagonalized as in Subsection \ref{subsec:ND_num}. In Section IV, we analyze the accuracy of this approximation for system setups for which $\mathbf{N}_\mathrm{RIS, Rx}^\dag \mathbf{N}_\mathrm{RIS,Tx} \ne \mathbf{I}_N$.

Given the matrix $\mathbf{\Phi}^\mathrm{FOC}$, the RIS-aided end-to-end channel boils down to a conventional multiple-input multiple-output channel with matrix $\mathbf{Z}^\mathrm{FOC} = \mathbf{Z}(\mathbf{\Phi}^\mathrm{FOC})$. Therefore, the transmitter can be optimized by applying the SVD to it, which yields $\mathbf{Z}^\mathrm{FOC} = \mathbf{U} \mathbf{S} \mathbf{V}^\dag$ with $\mathbf{S} = \diag \{\lambda_1, \lambda_2,..., \lambda_K\}$. Then, the SNR of the $k$-th stream is
\begin{equation}
    s_k^\mathrm{FOC-NUM} = \frac{|\lambda_k|^2}{\sigma^2}
\end{equation}

\begin{table}[!t]
\centering
\begin{tabular}{|c|c|c|}
\hline
Scheme  & RIS configuration & Input Covariance Matrix \\ \hline
ND-NUM  & $\mathbf{\Phi}^\mathrm{ND-NUM}$ \eqref{eq:RIS_ND_NUM} & $\mathbf{Q}^\mathrm{ND-NUM}$ \eqref{eq:Q_ND_NUM}\\ \hline
ND-PSWF  & $\mathbf{\Phi}^\mathrm{ND-PSWF}$ \eqref{eq:RIS_ND_PSWF} & $\mathbf{Q}^\mathrm{PSWF}$ \eqref{eq:Q_PSWF}\\ \hline
FOC-NUM & $\mathbf{\Phi}^\mathrm{FOC}$ \eqref{eq:RIS_FOC}& $\mathbf{Q}^\mathrm{FOC-NUM}$ \eqref{eq:Q_FOC_NUM}\\ \hline
FOC-PSWF & $\mathbf{\Phi}^\mathrm{FOC}$ \eqref{eq:RIS_FOC} & $\mathbf{Q}^\mathrm{PSWF}$ \eqref{eq:Q_PSWF}\\ \hline
\end{tabular}
\caption{Summary of the considered schemes. Legend: Non-diagonal RIS (ND), Diagonal RIS (FOC), Numerical computation (NUM), Closed-form expression (PSWF).}
\label{tab:summarySchemes}
\end{table}

The computation of the power allocation matrix $\mathbf{P}^\mathrm{FOC-NUM}$ is discussed in Subsection \ref{subsec:powerAlloc}. Therefore, the input covariance matrix is
\begin{equation}
\label{eq:Q_FOC_NUM}
    \mathbf{Q}^\mathrm{FOC-NUM} = \mathbf{V} \mathbf{P}^\mathrm{FOC-NUM}\mathbf{V}^\dag
\end{equation}

\subsection{Diagonal RIS Based on Analysis}
This scheme is obtained using the same line of thought as the scheme in Subsection III-B. Specifically, $\mathbf{\Phi}$ is set to $\mathbf{\Phi}^\mathrm{FOC}$ as in Subsection III-C, and the covariance matrix $\mathbf{Q}$ is obtained by sampling PSWFs as described in Subsection III-B. 

For completeness, the matrices $\mathbf{\Phi}$ and $\mathbf{Q}$ corresponding to the four schemes considered in the present paper are summarized in Table \ref{tab:summarySchemes}.

\subsection{Power Allocation}
\label{subsec:powerAlloc}
The performance of the four schemes considered in the present paper depends on the power allocated to the communication modes. The optimal performance is achieved by applying the water-filling algorithm \cite{Tse}.

More precisely, the matrix $\mathbf{P}$ is defined as
\begin{equation}
    \mathbf{P} = \diag \{P_1, P_2,...,P_K\}
\end{equation}
where $K = \min(L,M)$ and $P_k$ is given by
\begin{equation}
    P_k = \max\left(0, \mu - \frac{1}{s_k}\right)
\end{equation}
where $s_k$ is the SNR of $k$-th communication mode, as defined in the previous subsections, and $\mu$ is chosen to fulfill the condition $\sum_{i=1}^{K}P_i = P_T$.

\section{Numerical Examples}
\label{sec:NumResults}
In this section, we illustrate some numerical results to analyze and compare the performance of the four considered case studies. In the analyzed setup, the transmitter and receiver are equipped with $8 \times 8$ elements, and the RIS consists of $32 \times 32$ elements. Also, we set $l_t=5$ m, $d_t=5$ m, $f = 3.5$ GHz, $P_T = -20$ dBm, and $\sigma^2 = -97$ dBm. The four schemes considered in Subsections III-A, III-B, III-C, and III-D are denoted as ND-NUM, ND-PWSF, FOC-NUM, and FOC-PWSF, respectively.

\begin{figure}[!t]
    \centering
   \includegraphics[width=\columnwidth]{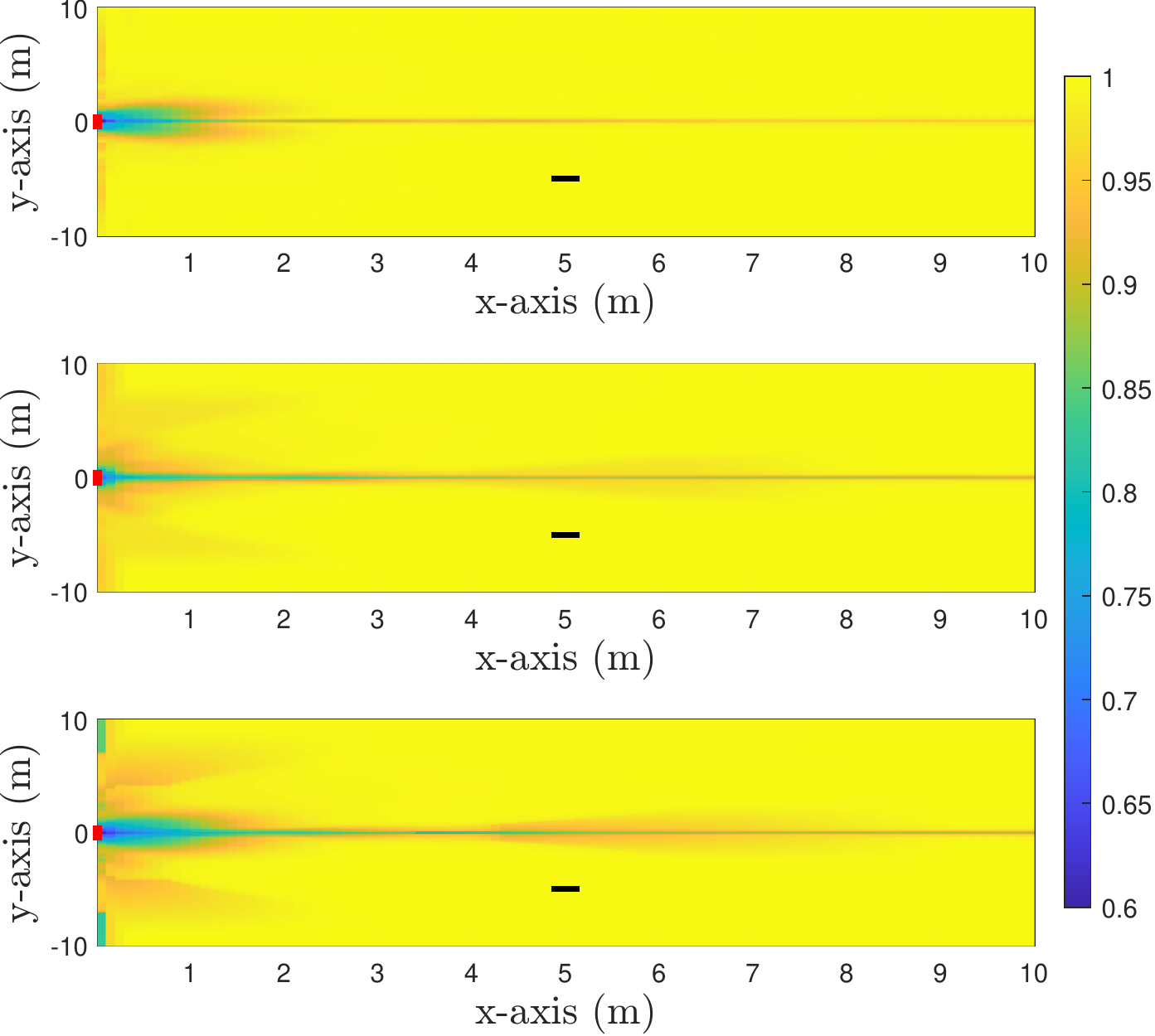}
    \caption{(top) Ratio of the achievable rates between ND-PWSF and ND-NUM. (center) Ratio of the achievable rates between FOC-NUM and ND-NUM. (bottom) Ratio of the achievable rates between FOC-PWSF and ND-NUM.}
    \label{fig:maps1} \vspace{-0.25cm}
\end{figure}

\begin{figure}[!t]
    \centering
   \includegraphics[width=\columnwidth]{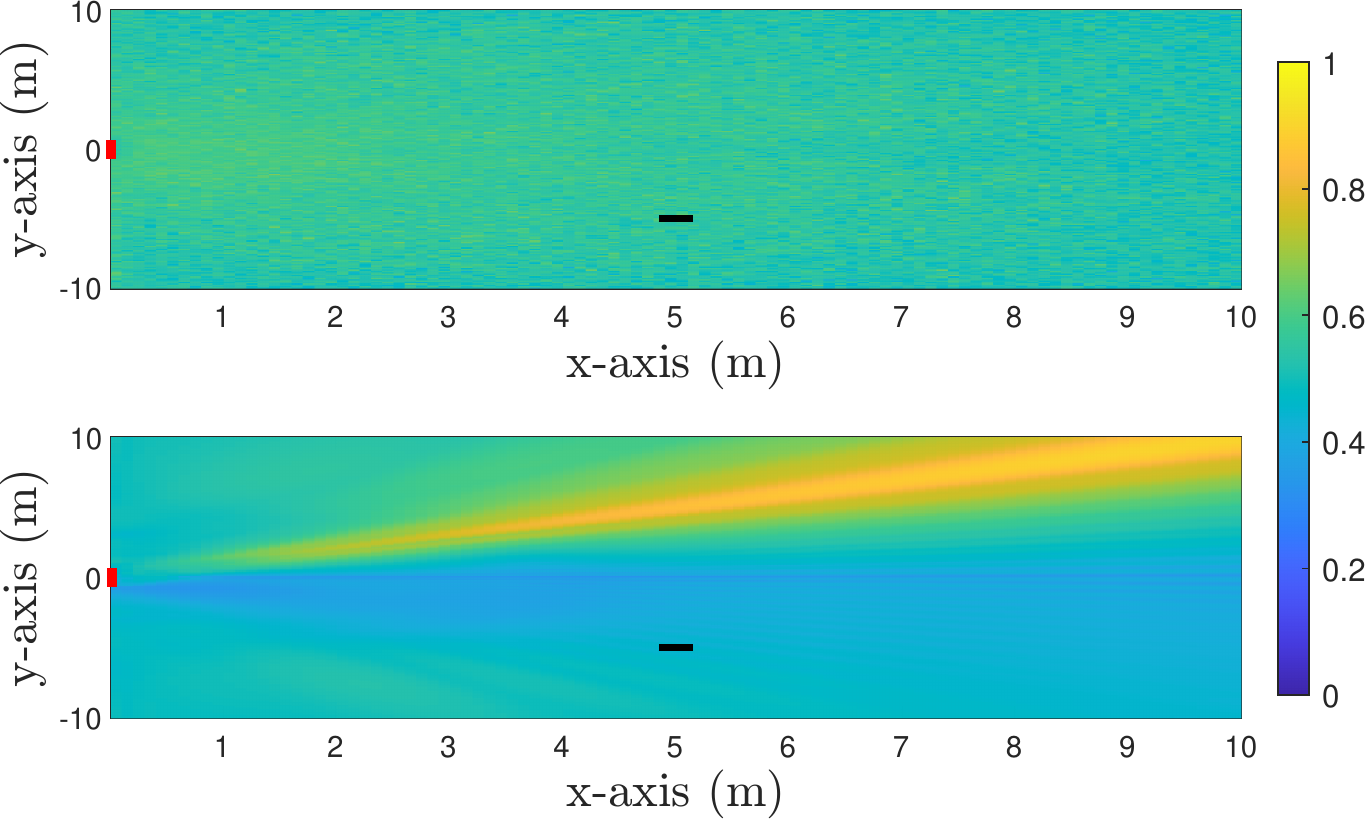}
    \caption{(top) Ratio of the achievable rates between a random RIS configuration scheme and ND-NUM. (bottom) Ratio of the achievable rates between a uniform (specular reflection) RIS configuration scheme and ND-NUM.}
    \label{fig:maps2} \vspace{-0.25cm}
\end{figure}

Figure \ref{fig:maps1} illustrates the ratio between the achievable rates of the schemes ND-PWSF, FOC-NUM, FOC-PWSF, and the benchmark scheme ND-NUM. The transmitter and the RIS are represented by black and red segments, respectively. The low complexity schemes ND-PWSF and FOC-PWSF provide rates similar to those obtained with the aid of numerical methods. Also, Fig. \ref{fig:maps2} confirms that an RIS that is configured as a random scatterer or as a specular reflector largely underperforms the considered schemes where the RIS is optimized. 

\begin{figure}[!t]
    \centering
   \includegraphics[width=0.9\columnwidth]{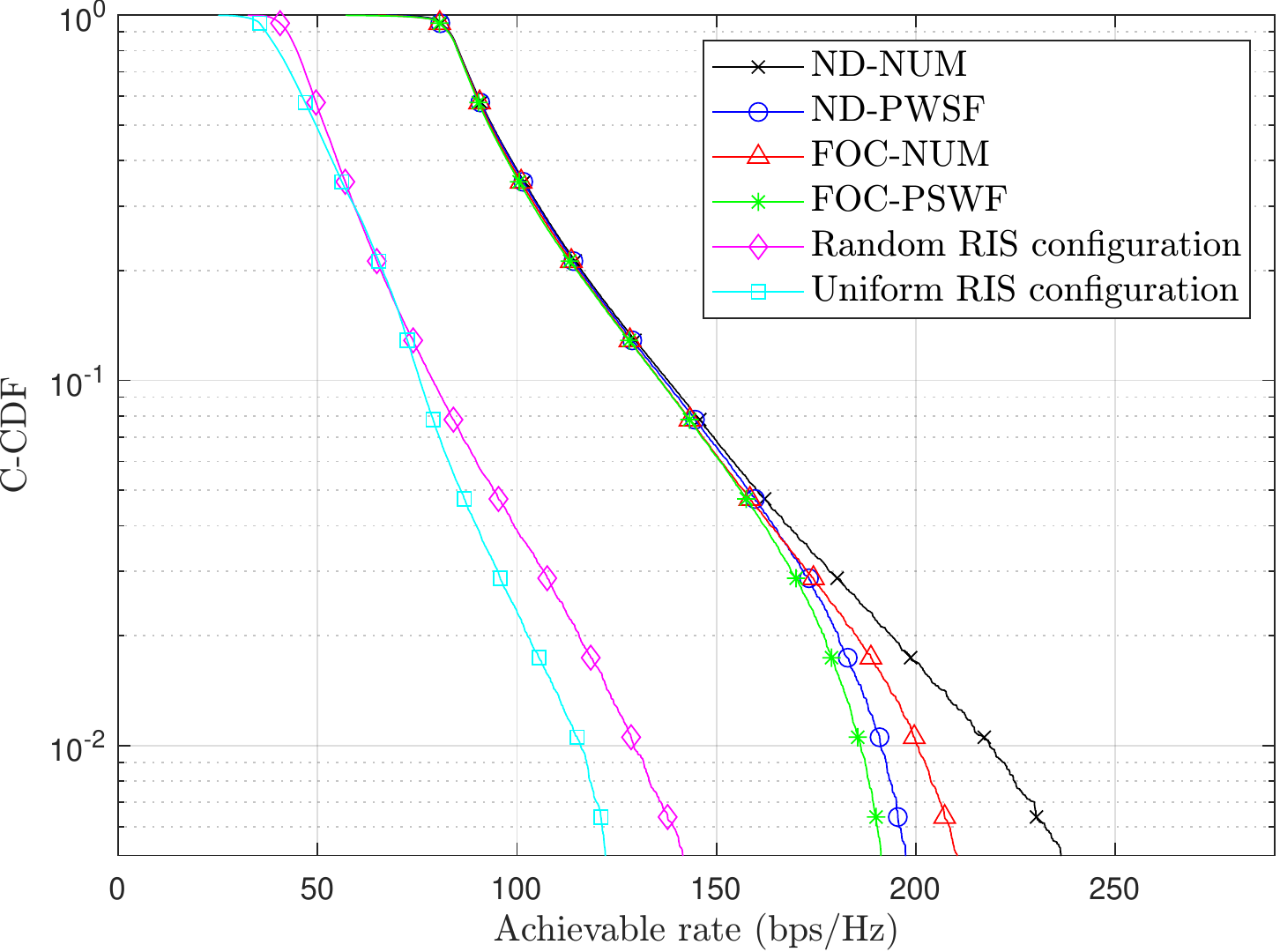}
    \caption{C-CDF of the achievable rates.}
    \label{fig:ccdf} \vspace{-0.25cm}
\end{figure}

Finally, Fig. \ref{fig:ccdf} shows the complementary cumulative distribution function (C-CDF) of the considered four schemes. The C-CDF confirms that the proposed closed-form expressions for the  RIS configuration and input covariance matrix offer rates that are very close to those obtained with computationally intensive numerical methods. Some differences are noticeable, in the considered case study, only for large values of the rates.

\section{Conclusions}
\label{sec:Conclusions}
We have analyzed low complexity designs for RIS-aided HoloS communication systems, in which the configuration of the HoloS transmitter and the RIS are given in a closed-form expression. We have considered implementations based on diagonal and non-diagonal RISs. Over line-of-sight channels, we have shown that the proposed schemes provide rates similar to those obtained through complex numerical methods.

\bibliographystyle{IEEEtran}
\bibliography{sample}

\end{document}